\documentclass[]{aastex63}

\usepackage{amsmath}
\usepackage{amsfonts}
\usepackage{amssymb}
\usepackage{upgreek}
\usepackage{graphics}
\usepackage{hyperref}

\begin{document}
%\shortauthors{Huston and Luhman}
%\shorttitle{IMF in W3}
\title{The Initial Mass Function of Low-mass Stars and Brown Dwarfs in the W3 Complex}
\author[0000-0003-4591-3201]{M. J. Huston}
\affiliation{Department of Astronomy and Astrophysics, The Pennsylvania
State University, University Park, PA 16802, USA}
\affiliation{Center for Exoplanets and Habitable Worlds, The
Pennsylvania State University, University Park, PA 16802, USA}
\affiliation{Penn State Extraterrestrial Intelligence Center, The Pennsylvania State University, State College, PA 16802, USA}

\author{K. L. Luhman}
\affiliation{Department of Astronomy and Astrophysics, The Pennsylvania
State University, University Park, PA 16802, USA}
\affiliation{Center for Exoplanets and Habitable Worlds, The
Pennsylvania State University, University Park, PA 16802, USA}

\begin{abstract}
We have used archival infrared images obtained with the Wide Field Camera 3 on board the Hubble Space Telescope to
constrain the initial mass function of low-mass stars and brown dwarfs in the W3 star-forming region.
The images cover 438~arcmin$^2$, which encompasses the entire complex, and were taken in the filters
F110W, F139M, and F160W. 
We have estimated extinctions for individual sources in these data from their colors and have
dereddened their photometry accordingly. By comparing an area of the images that contains the
richest concentration of previously identified W3 members to an area that has few members and is
dominated by background stars, we have estimated the luminosity function for members of W3 with masses of
0.03--0.4~$M_\odot$. That luminosity function closely resembles data in typical
nearby star-forming regions that have much smaller stellar populations than W3 ($\lesssim$500 vs. several thousand objects).
Thus, we do not find evidence of significant variations in the initial mass function of low-mass stars and brown dwarfs with star-forming conditions, which is consistent with recent studies of other distant massive
star-forming regions.
\end{abstract}

\section{Introduction} \label{sec:intro}
Measuring the Initial Mass Function (IMF) of stars and brown dwarfs is essential for constraining the star formation process. As reviewed by \cite{imf_review}, the high-mass IMF ($M \gtrsim 1$~$M_\odot$) in most populations is well-described by a power law with a slope of $\Gamma \sim 1.35$ \citep{salpeter} for $dN/d\log m \propto m^{-\Gamma}$. Below $\sim 1$~$M_\odot$, the IMF flattens, reaching a maximum near 0.2~$M_\odot$. Because low-mass stars and brown dwarfs are dim, their IMF has primarily
been studied in nearby stellar populations.
Based on surveys for members of the solar neighborhood ($<30$~pc) and the nearest young clusters and associations (50--300~pc, 1--100~Myr), the IMF declines from its peak into the the substellar regime ($M<0.08$~$M_\odot$) with a slope in the range $ -1 \lesssim \Gamma \lesssim 0$. \cite{imf_review} concluded that there is no clear evidence for large variations of the IMF with star-forming conditions. However, the low-mass IMF has been measured in a more limited
range of (local) environments, so there are fewer constraints on variations
in that mass range.
For instance, nearby clusters and associations typically contain a few hundred members (with a few OB stars), and one of the richest, the Orion Nebula Cluster, has $\sim$2000 members ($\sim$20 OB stars). To search for variations of the low-mass IMF in more extreme conditions,
\cite{rcw38,rosette} performed deep infrared (IR) imaging of RCW 38 and NGC 2244 ($\sim$1.6~kpc) down to 0.02~$M_\odot$. They derived IMFs with slopes of $\Gamma = -0.29\pm 0.11$ for 0.02--0.5~$M_\odot$ in RCW 38 and
$0.03\pm 0.02$ for 0.02--0.4~$M_\odot$ in NGC 2244, which are within the range of values found in nearby populations.

Like RCW 38 and NGC 2244, W3 is a rich star-forming complex in which one can search for variations in the low-mass IMF. It is located in the Perseus arm at a distance of $2.1 \pm 0.2$~kpc \citep{w3_gaia} and contains $\sim$100 OB stars \citep{kim15}. The three
largest stellar populations in W3 are the central W3 Cluster (also known as IC 1795),  W3 (OH), and W3 Main \citep{rom15}. Previous IMF measurements in W3 have been primarily confined to the stellar regime \citep{Bik_2014,kim15}. 

W3 is among the distant star-forming complexes imaged by the Hubble Space Telescope's (HST) 
Wide Field Camera 3 (WFC3) through program 15238 (A. Kraus). Those images offer the 
potential for identifying members of W3 down to very low masses
because of their depth, high angular resolution, and combination of filters,
where the latter are well-suited for distinguishing late-type W3 members from most
field stars.
In this work, we reduce the WFC3 images for W3 and measure photometry for all detected sources (Section \ref{sec:reduction}). We use the WFC3 colors to estimate the extinctions for 
individual objects and derreden their photometry accordingly. We estimate the background
star population toward W3 and apply a correction for it to derive the luminosity
function for low-mass stars and brown dwarfs in W3, which is compared to similar measurements 
in nearby star-forming regions to check for a variation in the IMF (Section~\ref{sec:analysis}).

\section{Observations and Data Reduction} \label{sec:reduction}
\subsection{Image Collection}
\begin{figure}
\centering
    \includegraphics[width=1\textwidth]{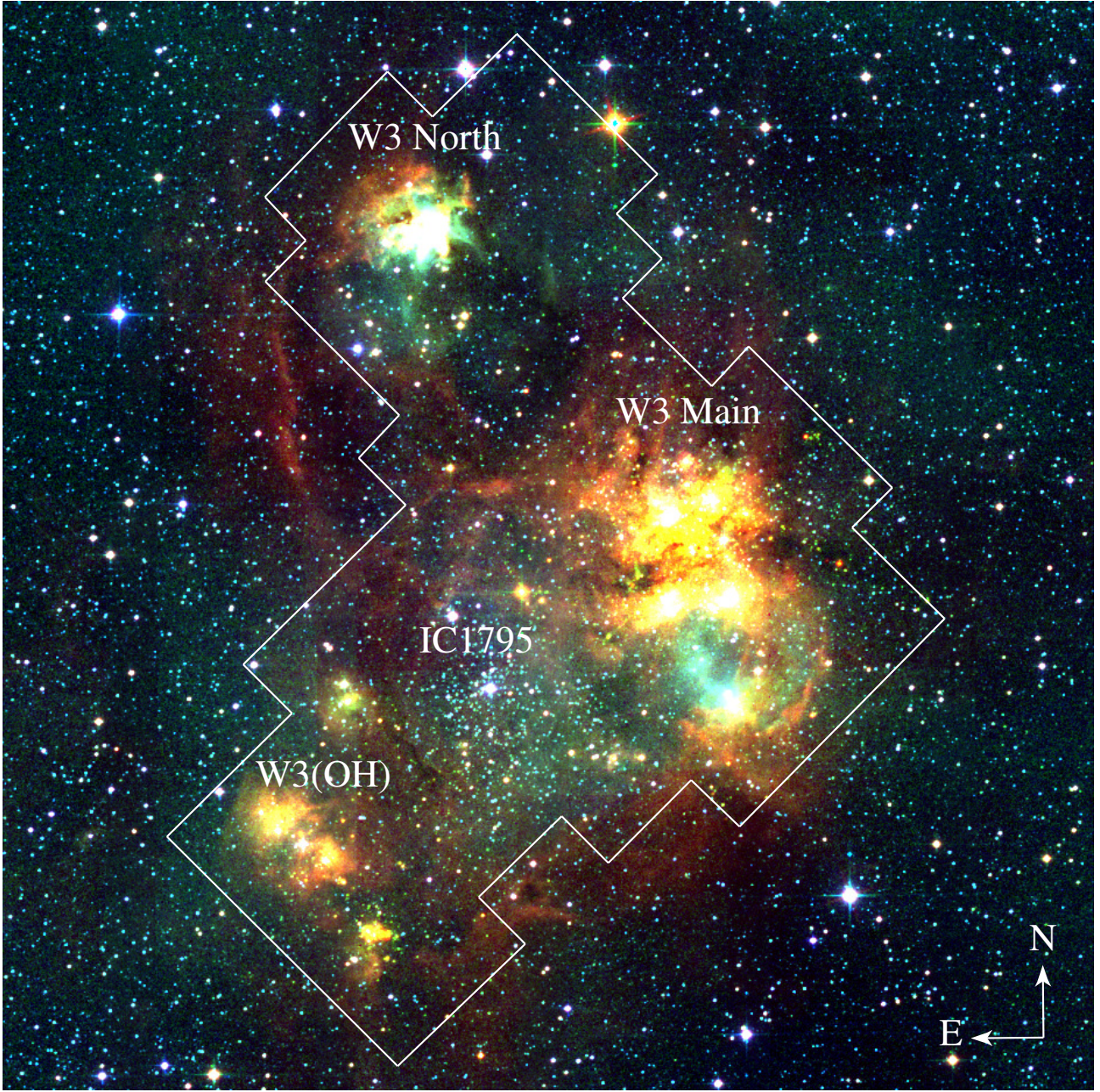}
    \caption{Images of the W3 complex at $J$, $K_s$, and W1 from 2MASS and WISE (blue, green, red), which cover an area of
    $0\fdg6\times0\fdg6$. The boundary of the field observed by WFC3 is marked.}
    \label{fullimage}
\end{figure}

Images of W3 were taken with the IR channel of WFC3 \citep{wfc3} through program 15238 (PI: A. Kraus). The camera's array has a format of $1014 \times 1014$ pixels where each pixel has a size of $\sim 0\arcsec.135 \times 0\arcsec.121$, corresponding to a field of view of $136\arcsec \times 123\arcsec$. WFC3 imaged 102 adjacent fields that overlapped slightly. The resulting mosaic covers an area of 438~arcmin$^2$. The boundary of the mosaic and the locations of the main components of the W3 complex are marked in Figure~\ref{fullimage}, which presents a false color image
constructed from data at $J$ and $K_s$ from the Two Micron All-Sky Survey \citep[2MASS,][]{skr06} and data at W1 (3.4~\micron) from
the Wide-field Infrared Survey Explorer \citep[WISE,][]{wri10}.

The images were obtained with the drift-and-shift (DASH) method \citep{dash}, which efficiently images large areas by performing a guide star acquisition only on the first pointing in an orbit and guiding with gyros alone for additional pointings. The IR channel uses the MULTIACCUM readout mode, sampling the signal multiple times during an exposure. Each exposure for W3 consisted of 4--14 reads, each of which was 25 or 50~s in length. When guiding with gyros alone, the telescope can drift noticeably during an exposure of a few minutes, but the drift is often less than one pixel within the 25/50~s reads. As a result, the individual reads can be shifted to correct for the telescope drift and combined to produce an unsmeared image.

The WFC3 images were taken with three filters, F110W (0.9--1.2~$\mu$m), F139M (1.35-1.41 $\mu$m), and F160W (1.4--1.7 $\mu$m). 
The F139M filter measures absorption bands
in H$_2$O and CH$_4$ that appear in late-type objects while the other two filters measure the adjacent continuum. Those bands are more difficult to observe with ground-based telescopes since they occur at
wavelengths where telluric absorption is very strong.
The resulting colors from this set of filters can help discriminate between low-mass members of W3 and field stars, even in the presence of reddening.

\subsection{Image Reduction}
The raw WFC3/IR images of W3 were retrieved from the Mikulski Archive for Space Telescopes\footnote{\url{http://archive.stsci.edu/}}. 
Our data reduction procedures are similar to those applied to WFC3 images of IC 348 taken with the DASH method \citep{ic348_LH}.
We used the python routine {\tt wfc3dash}\footnote{\url{https://github.com/gbrammer/wfc3dash}} to split each raw image into its constituent 25/50~s reads, which were aligned and combined with the tasks {\tt tweakreg} and {\tt astrodrizzle} in the Drizzlepac package\footnote{\url{https://github.com/spacetelescope/drizzlepac}}. The resulting images were given a resampled plate scale of $0\farcs065$~pixel$^{-1}$. We aligned the world coordinate system of each image to the astrometry of sources from the second data release of the Gaia mission \citep{gaiadr2} that were detected and unsaturated in the WFC3 images.

For each image, we identified the detected sources and measured their pixel coordinates using the {\tt starfind} routine in IRAF. Detection thresholds were selected to be low
enough that the routine identified most sources that are visually apparent in areas of low background emission. As a result, some spurious detections were identified in areas with brighter background levels. A spurious detection in one filter is unlikely to have counterparts at the same location in the other filters, so sources appearing in only one filter were rejected when the catalogs for the three bands were merged. Photometry was measured for each source using {\tt phot} in IRAF with an aperture radius of four pixels and radii of four and eight pixels for the inner and outer boundaries of the sky annulus, respectively.

The photometric calibration for WFC3 is defined for a radius of $0\farcs4$, which corresponds to 6.15
pixels in our reduced images. As a result, it was necessary to apply an aperture correction to the
instrumental magnitudes measured with {\tt phot}.
Due to the variable drift rate when guiding only with gyros, the sampling of the point spread function, and hence the aperture correction, can vary among the reduced images for a given filter. Therefore, we measured an aperture correction for each reduced image, which was calculated from photometry for bright, unsaturated sources. We applied the corrections and the zero-point Vega magnitudes of 25.74 (F110W), 23.05 (F139M), and 24.30 (F160W) for $0\farcs4$ to the photometry. The average values of the aperture corrections were 0.14, 0.16, 0.20~mag for F110W, F139M, and F160W, respectively.

Due to the small overlap between neighboring images in the WFC3 mosaic, some sources were detected in multiple images. We used the Starlink Tables Infrastructure Library Tool Set \cite[STILTS,][]{stilts} to identify duplicate detections in a given band and we adopted their mean photometry and astrometry. We then used STILTS to match the catalogs for the three filters to produce a single final catalog. As mentioned earlier, sources that appeared in only a single band were rejected. The resulting catalog contains 38,708 sources and is presented in Table~\ref{tabl}.

\begin{deluxetable}{ll}
    \tabletypesize{\scriptsize}
    \tablewidth{0pt}
    \tablecaption{Sources in WFC3 Images of W3\label{tabl}}
    \tablehead{
    \colhead{Column Label} &
    \colhead{Description}}
    \startdata
    RAdeg & R.A. (J2000)\\
    DEdeg & Decl. (J2000) \\
    110mag & F110W magnitude from HST WFC3 \\
    e\_110mag & Error in 110mag \\ 
    139mag & F139M magnitude from HST WFC3 \\
    e\_139mag & Error in 139mag \\ 
    160mag & F160W magnitude from HST WFC3 \\
    e\_160mag & Error in 160mag 
    \enddata
    \tablecomments{This table is available in its entirety in a machine-readable form.}
\end{deluxetable}

\section{Analysis}\label{sec:analysis}
\subsection{Color-color and Color-magnitude Diagrams}

\begin{figure}
\centering
    \includegraphics[width=\textwidth]{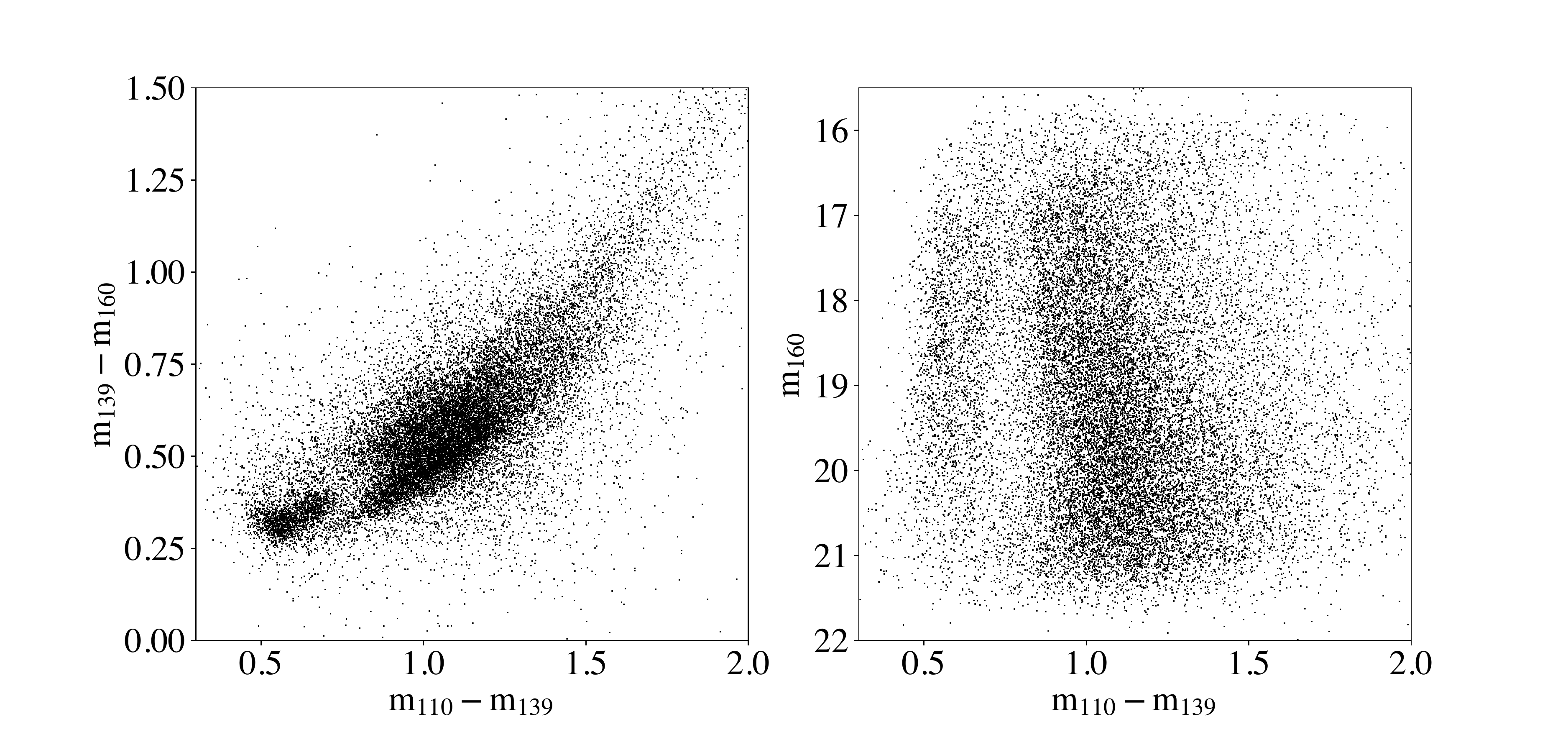}
    \caption{Color-color and color-magnitude for sources in the WFC3 images of W3 that have errors of $<0.1$~mag in F110W, F139M, and F160W. The blue and red populations likely correspond to field dwarfs and a mixture of W3 members and background
    giants, respectively.}
    \label{colmag}
\end{figure}

To analyze the WFC3 photometry in W3, we begin by plotting the sources with errors of 
$<$0.1~mag in all three bands in
color-color and color-magnitude diagrams in Figure~\ref{colmag}. As illustrated in the color-magnitude diagram, the saturation and detection limits in that sample
are near $m_{160}\sim$16 and 21.5, respectively. Because one of the filters (F139M) is
much narrower than the other two, the three-band sample is limited by the shallower depth 
of that filter. Sources detected in F110W and F160W but not F139M extend down to 
$m_{160}\sim$23.
We estimate a completeness limit of $\sim$21.5 in F160W based on the magnitude at
which the number of sources begins to flatten and turn over. These data are deeper
than the previously available images of W3 at similar wavelengths \citep{rom15}.
The fraction of stars with detections in three bands drops below 90\% at
$m_{160}\gtrsim$20.8.
The two diagrams in Figure~\ref{colmag} exhibit distinct blue and red populations, which 
likely correspond to field dwarfs and a mixture of cluster members and background giants, 
respectively. Similar populations have been detected in previous near-IR imaging of W3 
\citep{Bik_2014}.
In the color-color diagram, the red population appears to contain upper and lower components, which are likely low-mass
W3 members and background giants, respectively, based on the analysis in Section~3.3.

\subsection{Defining Cluster and Background Fields}

\begin{figure}
\centering
    \includegraphics[width=0.6\textwidth]{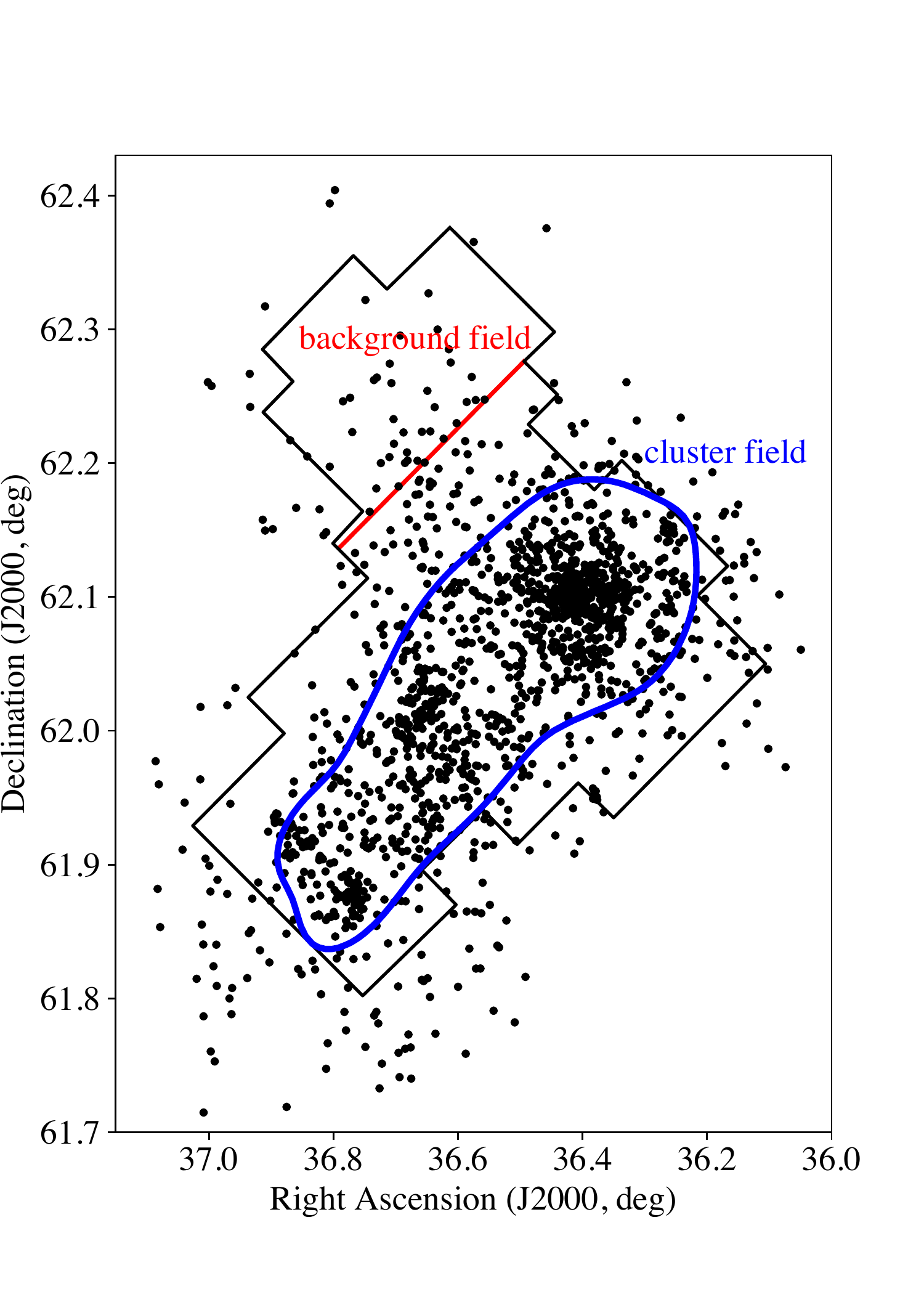}
    \caption{Map of the boundary of the field toward the W3 complex imaged by WFC3 (black line) and candidate members previously identified with X-ray data \citep{mpcm}. For our analysis, we have selected a density contour for those candidates to define a cluster field (blue line) and we have selected a section of the WFC3 field that has few candidates for measuring the background star population (north of the red line).}
    \label{spatial}
\end{figure}

The red population in our color-color diagram is likely a mixture of W3 members and background stars. Therefore, characterizing the IMF of W3 requires
a correction for the contribution of background stars. We begin by defining one field that should have a high concentration of cluster members and another field that should have few
cluster members so that it can measure the background population. To do that, we make
use of the candidate stellar members of W3 identified with X-ray data by \cite{mpcm}.
The spatial distribution of those candidates is shown with the boundary of the WFC3
field in Figure~\ref{spatial}. We have derived density contours for the X-ray candidates
using a Gaussian kernel-density estimator, and we have selected one
that encompasses most ($\sim$75\%) of the X-ray candidates as the boundary of
the cluster field for our analysis. This field should be large enough that our
IMF results are not affected by mass segregation. The vicinity of the W3 North HII region contains few X-ray candidates \citep{feigelson}, so we have selected that section
of the WFC3 field for characterizing the background star population.

\subsection{Extinction Estimates}

\begin{figure}
\centering
    \includegraphics[width=\textwidth]{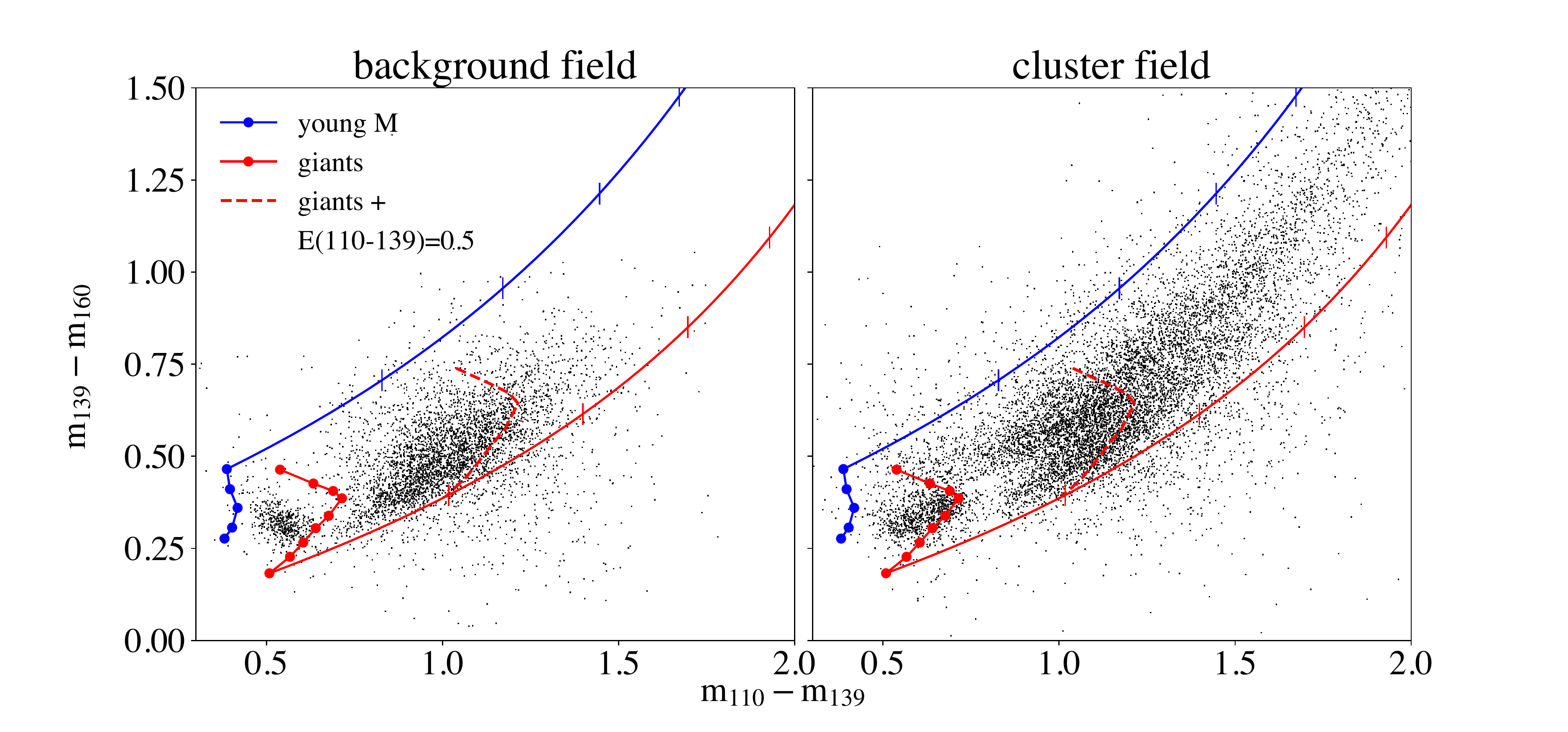}
    \caption{Color-color diagrams for WFC3 sources in the cluster and background fields defined in Figure \ref{spatial}. We have marked the intrinsic colors of young M5--M9 objects \citep[blue points, bottom to top,][]{mdwarf_standard} and K/M giants \citep[red points,][]{giantmodels}. Reddening vectors are indicated for the latest young object and the earliest giant based on the extinction law of \citet{schlafly_2016} with ticks at $A_{160}= 1$, 2, etc.
    We also show the giant sequence with a reddening of $E(110-139)=0.5$ (dashed lines), which is a threshold used for selecting stars for
    the luminosity functions in Figure~\ref{hists}.
    }
    \label{colcol}
\end{figure}

In Figure~\ref{colcol}, we have plotted color-color diagrams for sources
in the cluster and background fields.
As in Figure~\ref{colmag}, two distinct populations are apparent in each diagram, which
we attribute to field dwarfs (bluer) and a combination of W3 members and background
giants (redder).
Both the blue and red populations in the cluster field extend to redder colors
than the two populations in the background field, indicating that the extinctions
in the cluster field extend to higher values and that the blue population includes
both foreground and background dwarfs.
In addition, the red population in the cluster field contains a prominent upper component that is absent from the red population in the background field, which presumably
consists of low-mass members of W3.
The blue populations of field dwarfs in the background and fields are omitted
from all remaining analysis.

We have simulated the intrinsic WFC3 colors of low-mass W3 members using
the standard spectra of young M5--M9 objects from \citet{mdwarf_standard} and
we have simulated the colors of giants using the model spectra from 
\citet{giantmodels} for temperatures of 2900--4400~K (K and M types).
The resulting colors are shown in both color-color diagrams in Figure~\ref{colcol}. 
In addition, to illustrate the effect of extinction on the WFC3 colors,
we have plotted the reddening vectors for a young M9 object
and a 2900~K giant based on the extinction law of \citet{schlafly_2016}.
For each source in the color-color diagrams in Figure~\ref{colcol}, we have derived
an extinction by dereddening it to the sequence of intrinsic colors for giants (if
it intersects with that sequence). We also have estimated a second set of extinctions
assuming the intrinsic colors of the young M stars.
The latter values of $A_{160}$ are larger than those based on giant colors
by an average of $\sim0.5$~mag.

\subsection{F160W Luminosity Function for W3 Members}

As often done in measurements of the IMF in star-forming regions 
\citep[e.g.,][]{ic348,taurus}, we have attempted to constrain the IMF in W3
using an extinction-limited sample of stars. We have selected a
limit that is high enough to encompass a large number of stars but 
low enough that the completeness limit reaches fairly low masses, arriving at a 
compromise of $E(110-139)<0.5$ (when based on giant colors), which corresponds to
$A_{160}\lesssim1$. We apply that limit when selecting stars in both the background
and cluster fields.

\begin{figure}
\centering
    \includegraphics[width=0.6\textwidth]{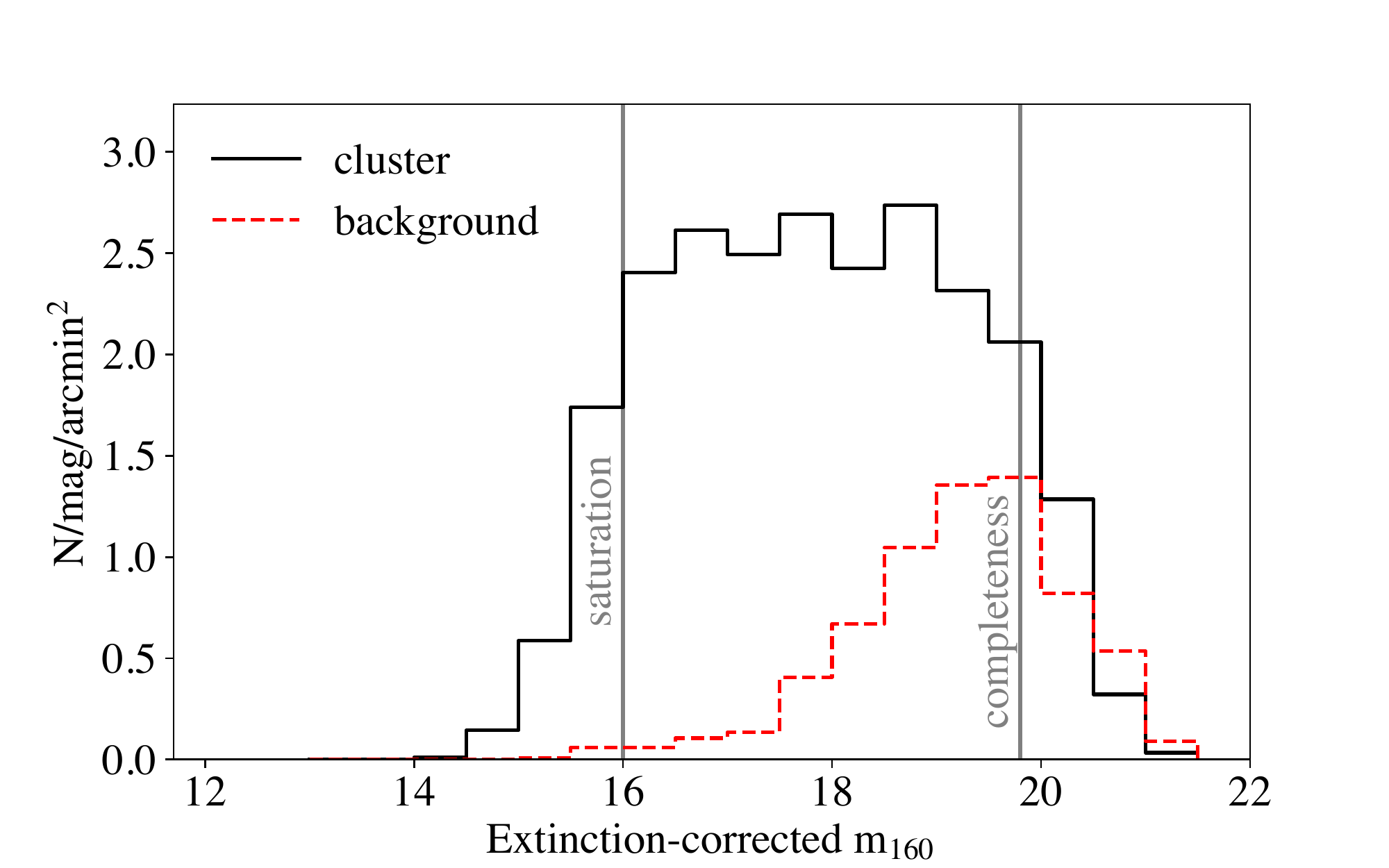}
    \caption{F160W luminosity functions for objects with extinction-corrected colors of $m_{139}-m_{160}>0.4$ and $E(110-139)<0.5$ (relative to giant
    colors) in the cluster field (black solid line) and background field (red dashed line).}
    \label{hists}
\end{figure}

We seek to characterize the low-mass IMF in W3 in terms of 
F160W luminosity function (LF), which we define as the 
histogram of the number of stars per area as function of 
extinction-corrected $m_{160}$.
To estimate the LF for W3, we need to use the data in the background field
to derive a correction for the background contamination in the cluster field.
For the background field, we have dereddened the photometry of the stars
with $E(110-139)<0.5$ using the extinctions derived in the previous
section assuming the intrinsic colors of giants and we have constructed a LF
with those dereddened data.
Since extinction reaches higher levels in the cluster field than in the background 
field, the surface density of background stars within a given extinction limit
are not the same in the two fields.
To account for that, we have scaled the LF from the
background field by a factor (0.37) such that it matches the LF of
stars in the cluster field that have $E(110-139)<0.5$ and that fall within a range of 
extinction-corrected colors where background stars should dominate within the
cluster field, namely $m_{139}-m_{160}=0.2$--0.35. That range of colors was
selected because it corresponds to sufficiently early spectral types (K and early M)
that W3 members should be saturated in the WFC3 images, and hence absent from our catalog.
At redder values of extinction-corrected $m_{139}-m_{160}$ (later spectral types),
W3 members are less likely to be saturated and they become increasingly dominant
over the background stars, which is evident from a visual comparison of the
color-color diagrams for the background and cluster fields
in Figure~\ref{colcol}. As an additional comparison that is more quantitative,
we show in Figure~\ref{hists} the LF for the cluster field and the scaled LF for the
background  field, both for extinction-corrected colors of $m_{139}-m_{160}>0.4$.
The former exhibits a clear excess of objects relative to the latter, which
corresponds to low-mass members
of W3. In Figure~\ref{hists}, we have marked the completeness limit in
extinction-corrected $m_{160}$ for the LF in the cluster field, which corresponds to
the magnitude above which most F160W sources have detections in all three bands
(cited earlier as $m_{160}\sim20.8$) dereddened by our adopted extinction limit
($A_{160}\sim1$).

\begin{figure}
\centering
    \includegraphics[width=\textwidth]{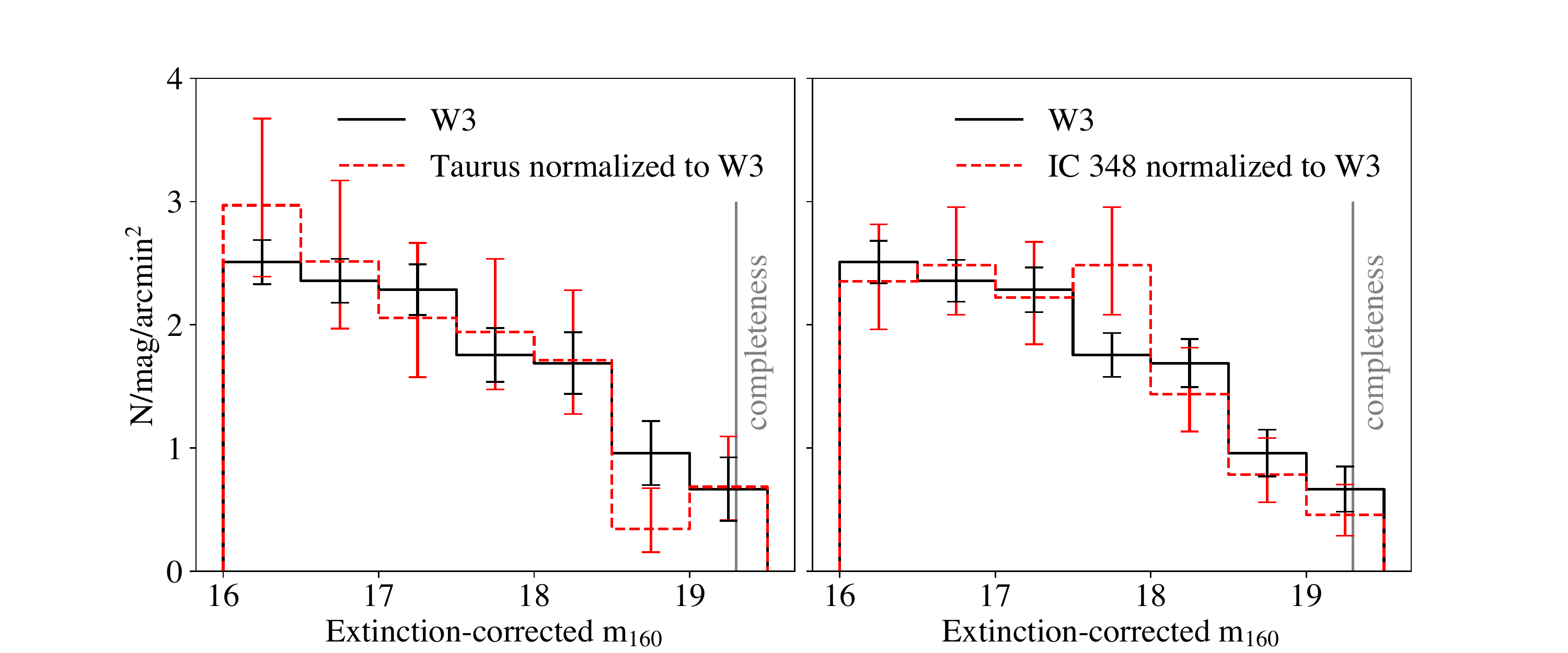}
    \caption{F160W luminosity functions for W3 members, calculated as the difference between the two distributions in Figure \ref{hists} (black
    solid histogram), and the distributions for two nearby star-forming
    regions adjusted to the distance of W3 \cite[red dashed histograms,][]{taurus,ic348}.}
    \label{lumfnc}
\end{figure}

We have computed the difference of the two LFs in Figure~\ref{hists} to
estimate the LF of low-mass members of W3. In the preceding analysis, 
the extinctions used to deredden the photometry in those LFs
were based on the intrinsic colors of giants, but low-mass and brown dwarfs stars
in W3 should have bluer intrinsic colors than giants (Figure~\ref{colcol}).
Therefore, we have shifted the LF 
for W3 to brighter magnitudes by 0.5~mag to account for the higher extinctions that
one would derive using the bluer intrinsic colors of young M-type objects.
Figure~\ref{lumfnc} shows the resulting LF for W3,
which is compared to LFs for two nearby star-forming regions, Taurus and IC~348,
adjusted to the distance of W3. For the latter two regions, we have adopted
extinction-limited samples of members from \citet{taurus} and \citet{ic348},
which are relatively large and have completeness limits that reach 
very low masses ($\sim0.01$~$M_\odot$).
Since F160W overlaps with the $H$ band, we have 
used the young standard spectra from \citet{mdwarf_standard} and the
filter profiles for $H$ and F160W to estimate the typical value of $m_{160}-H$ for
young low-mass stars and brown dwarfs and we have used it to convert the $H$-band data
in Taurus and IC~348 to $m_{160}$.

The LF of a young stellar population is a product of both its IMF and its age.
An age difference between W3 and Taurus or IC~348 would manifest as an offset between
their LFs that is roughly uniform across the magnitude range in question.
For instance, evolutionary models predict that the luminosities of low-mass stars
decrease by $\sim1$~mag between 1 and 3 Myr \citep{models}.
W3 lacks the extensive spectroscopy of its members that would be necessary for 
determining whether it has the same age as Taurus or IC~348, but 
they are probably coeval to within a few Myr
given that all three populations are still associated with molecular clouds
and have substantial disk fractions \citep{rom15,taurus,ic348}.
In that case, the similarity of W3's LF with the LFs of Taurus and IC~348 
(Figure~\ref{lumfnc}) would indicate that W3 shares a similar low-mass IMF with those 
regions. The limits of $m_{160}=16$ and 20 for the
LFs in Figure~\ref{lumfnc} correspond to median spectral types of $\sim$M3 and M8
based on the spectroscopic data in IC~348 and Taurus and masses of $\sim$0.4 and 
0.03~$M_\odot$ based on evolutionary models for ages of a few Myr \citep{models}.
In theory, the IMF in W3 could be measured down to the limit of $m_{160}\sim22$ for 
three-band detections, which corresponds to masses of $\gtrsim0.01$~$M_\odot$ for
the extinction range exhibited by most W3 members ($A_{160}>0.5$, 
Figure~\ref{colcol}). However, at those faintest magnitudes,
the LF of the cluster field does not show a significant excess relative
to the LF of the background field (Figure~\ref{hists}), and the resulting statistical
constraints on the LF for W3 are not useful.

\section{Conclusions}\label{sec:concl}
We have presented deep IR imaging of the W3 star-forming complex obtained with WFC3 on
HST. The images were taken with one filter that coincides with H$_2$O and CH$_4$ 
absorption bands (F139M) and two filters that cover the neighboring continuum
(F110W, F160W).
Using candidate members of W3 detected in previous X-ray imaging, we have identified
an area of the WFC3 field that should contain few W3 members so that we could estimate
the F160W LF for background stars. That LF has been subtracted from the data in
an area with the richest concentration of members to estimate the LF of
low-mass members of W3. The resulting LF extends from $m_{160}=16$--20, which
corresponds to masses of $\sim$0.03--0.4~$M_\odot$ based on evolutionary models for
ages of a few Myr. We find that the LF of low-mass W3 members is similar to the LFs in 
two examples of typical nearby star-forming regions, Taurus and IC~348.
Given that W3 has a much larger stellar population than nearby regions
($\lesssim$500 vs. several thousand), these results suggest that the low-mass IMF does 
not depend significantly on star-forming conditions, which
is consistent with recent studies of other distant massive star-forming regions 
\citep{rosette,rcw38}.

\section*{}
This work was supported by NASA grant 80NSSC18K0444 and is based on observations made with the NASA/ESA Hubble Space Telescope, obtained from the data archive at the Space Telescope Science Institute. STScI is operated by the Association of Universities for Research in Astronomy, Inc. under NASA contract NAS 5-26555. 
We thank the PI of HST program 15238, Adam Kraus, for planning those observations. The Center for Exoplanets and Habitable Worlds and the Penn State Extraterrestrial Intelligence Center are supported by the Pennsylvania State University and the Eberly College of Science. 
This work has made use of data from the European Space Agency (ESA) mission Gaia (\url{https://www.cosmos.esa.int/gaia}), processed by the Gaia Data Processing and Analysis Consortium (DPAC, \url{https://www.cosmos.esa.int/web/gaia/dpac/consortium}). Funding for the DPAC has been provided by national institutions, in particular the institutions participating in the Gaia Multilateral Agreement.
This research has made use of NASA’s Astrophysics Data System.
IRAF is distributed by the National Optical Astronomy Observatories, which is operated by the Association of Universities for Research in Astronomy, Inc. (AURA) under cooperative agreement with the National Science Foundation.

\bibliography{ms.bib}
\bibliographystyle{aasjournal}

\end{document}